# Ising Universality in Three Dimensions: A Monte Carlo Study

Henk W.J. Blöte, Erik Luijten and Jouke R. Heringa
Laboratorium voor Technische Natuurkunde, Technische Universiteit Delft,
P.O. Box 5046, 2600 GA Delft, The Netherlands

September 4, 1995

**Abstract**

We investigate three Ising models on the simple cubic lattice by means of Monte Carlo methods and finite-size scaling. These models are the spin-$\frac{1}{2}$ Ising model with nearest-neighbor interactions, a spin-$\frac{1}{2}$ model with nearest-neighbor and third-neighbor interactions, and a spin-1 model with nearest-neighbor interactions. The results are in accurate agreement with the hypothesis of universality. Analysis of the finite-size scaling behavior reveals corrections beyond those caused by the leading irrelevant scaling field. We find that the correction-to-scaling amplitudes are strongly dependent on the introduction of further-neighbor interactions or a third spin state. In a spin-1 Ising model, these corrections appear to be very small. This is very helpful for the determination of the universal constants of the Ising model. The renormalization exponents of the Ising model are determined as $y_t = 1.587$ (2), $y_h = 2.4815$ (15) and $y_i = -0.82$ (6). The universal ratio $Q = \langle m^2 \rangle^2 / \langle m^4 \rangle$ is equal to 0.6233 (4) for periodic systems with cubic symmetry. The critical point of the nearest-neighbor spin-$\frac{1}{2}$ model is $K_c = 0.2216546$ (10).

Keywords: *Phase transitions, Ising model, Monte Carlo*

PACS: 64.60.Ak, 64.60.Cn, 64.60.Fr, 75.10.Hk, 05.50.+q

## 1 Introduction

According to insights such as the universality hypothesis, the nature of a phase transition does not depend on the microscopic details of a system but only on global properties such as dimensionality and symmetry of the order parameter. Thus, it is believed that most three-dimensional systems with short-range interactions and a scalar order parameter (such as density or unidirectional magnetization) belong to the Ising universality class. This implies that the critical exponents, as well as other universal quantities, are identical for all these models. This universality class comprises, in addition to anisotropic magnetic systems, also models for alloys, gas–liquid systems and liquid mixtures.

In the case of two-dimensional Isinglike models, the evidence that universality holds is very strong. However, in three dimensions, where exact results are scarce and numerical techniques tend to be less accurate than in two dimensions, the situation is less satisfactory. Numerical uncertainties in the renormalization exponents amount to the order of several times $10^{-3}$. Since many years the most accurate results are those obtained by $\varepsilon$-, coupling-constant and series expansions [1–9], whereas recently quite accurate estimates have been obtained by the coherent-anomaly method as well [10]. However, new possibilities to investigate Isinglike models are now arising in parallel with the availability of fast and relatively cheap computers. While many systems in the supposed Ising class may be simulated with the help of these, spin models offer a clear advantage, at least as far as a study of the universal properties is concerned. This is because of the ease and efficiency of the Monte Carlo method, in particular of cluster algorithms. Thus, results from Monte Carlo-based methods [11–20] tend to become increasingly accurate.

However, slight differences occur between recent results for the scaling dimensions. One possible explanation is that universality is not satisfied. In order to solve the issue whether these deviations are real, it is desirable to obtain more accurate Monte Carlo data for the supposed universal quantities.

One problem that poses an obstacle to higher accuracies of these analyses is the presence of corrections to scaling. The dominant correction is attributed to an irrelevant renormalization exponent with an approximate value $y_i \simeq -0.83$ [7]. This means that the corrections decay relatively slowly and thus jeopardize the accuracy of the analysis. For this reason, we explore what modifications of the simple cubic Ising model with nearest-neighbor interactions can influence the amplitude of these corrections to scaling. If we can, in this way, suppress the irrelevant field, we may expect a decrease of the ill effects due to the corrections to scaling. One can, for instance, choose a different lattice structure. Series expansions using the body-centered cubic lattice [7] indicate that corrections to scaling are relatively small. However, here we prefer to introduce continuously variable parameters to adjust the irrelevant scaling field.

It is known [21] that the introduction of positive couplings with a range beyond the nearest neighbors in the simple cubic Ising model leads to a decrease of the correction-to-scaling amplitudes. We quote some preliminary results for the Hamiltonian

$$\mathcal{H}/k_\mathrm{B}T = -K_\mathrm{nn} \sum_{\langle \mathrm{nn} \rangle} s_i s_j - K_\mathrm{2n} \sum_{(\mathrm{2n})} s_i s_j - K_\mathrm{3n} \sum_{[\mathrm{3n}]} s_i s_j - K_\square \sum_\square s_i s_j s_k s_l \; , \tag{1}$$

where $\langle \mathrm{nn} \rangle$ indicates a sum over nearest-neighbor pairs, (2n) over second-neighbor pairs (diagonals of the elementary faces), [3n] over third-neighbor pairs (body diagonals in the elementary cubes) and $\square$ over four-spin products in all elementary faces of the cubic lattice. The associated couplings are denoted $K_\mathrm{nn}$, $K_\mathrm{2n}$, $K_\mathrm{3n}$ and $K_\square$, respectively. The spins $s_i$ can assume the values $+1$ and $-1$. These results were obtained by Monte Carlo simulation on the Delft Ising System Processor [22, 23]; they indicated that the introduction of positive $K_\mathrm{2n}$, $K_\mathrm{3n}$, or $K_\square$ reduces the correction-to-scaling amplitude. Third-neighbor couplings $K_\mathrm{3n}$ appear to be quite effective; for a ratio $K_\mathrm{3n}/K_\mathrm{nn} \approx 0.4$ the corrections become small. Much stronger second-neighbor couplings are required to obtain a similar effect [21].

Another approach is to introduce a third spin state $s_i = 0$: the spin-1 Ising model. The weight of the $s_i = 0$ state can be varied by means of a term $D \sum_i s_i^2$ in the Hamiltonian. Preliminary calculations showed that the corrections become small for $D \approx 0.7$. In our actual simulations we have used $D = \ln 2$, for reasons that will be explained in Sec. 2.

Thus, we have selected the following three Ising models: the spin-$\frac{1}{2}$ Ising model with $K_\mathrm{3n}/K_\mathrm{nn} = 0$ (the nearest-neighbor model), with $K_\mathrm{3n}/K_\mathrm{nn} = 0.4$, and the spin-1 model with $D = \ln 2$ and nearest-neighbor interactions. The algorithms used to simulate these models are described in Sec. 2. In addition to corrections to scaling, another obstacle to higher accuracies is the requirement of sufficiently accurate random numbers, in order to avoid biased results. In Sec. 3 we comment on the quality of our random-number generators and we mention consistency checks to which the algorithms were subjected. An analysis of the results for the dimensionless ratio $Q = \langle m^2 \rangle^2 / \langle m^4 \rangle$ is given in Sec. 4, followed by an analysis of the magnetic and temperature renormalization exponents in Sec. 5. The results for the three models satisfy universality: they are equal within the statistical inaccuracies. Assuming universality, $Q$ as well as the critical points of the three models can be obtained with a better precision, as is demonstrated in Sec. 6. Finally, a discussion of these results in relation with the existing literature and with fundamental questions concerning universality is presented in Sec. 7. As an appendix, we present experimental results for critical exponents of a number of phase transitions that have been supposed to belong to the 3-d Ising universality class.

## 2  Models and algorithms

The present Monte Carlo analysis concerns three different Ising models. These can be represented in terms of a spin-1 Hamiltonian on the simple cubic lattice:

$$\mathcal{H}/k_\mathrm{B}T = -K_\mathrm{nn} \sum_{\langle ij \rangle} s_i s_j - K_\mathrm{3n} \sum_{[kl]} s_k s_l + D \sum_m s_m^2 \; , \tag{2}$$

Table 1: The ratio $K_{3\mathrm{n}}/K_{\mathrm{nn}}$ and the value for $D$ (see Eq. (2)) for the three models.

| model # | $K_{3\mathrm{n}}/K_{\mathrm{nn}}$ | $D$ | description of model |
|---|---|---|---|
| 1 | 0 | $-\infty$ | spin-$\frac{1}{2}$ model with nn couplings |
| 2 | 0.4 | $-\infty$ | spin-$\frac{1}{2}$ model with nn and 3n couplings |
| 3 | 0 | $\ln 2$ | spin-1 model with nn couplings |

where nn and $\langle ij \rangle$ refer to nearest neighbors, and 3n and $[kl]$ to third-nearest neighbors (along body diagonals of the elementary cubes). The spins can assume three discrete values $s_i = 0, \pm 1$. The three models are specified in Table 1.

For $D = -\infty$ the $s_i = 0$ states are excluded and thus models 1 and 2 can be simulated by the Swendsen–Wang (SW) [24], the largest-cluster (LC) [25] or the Wolff [26] method. In cluster algorithms, one has to 'activate' a bond between two spins $s_i$ and $s_j$, coupled with strength $K_{ij}$, with a probability $p(K_{ij})\delta_{s_i,s_j}$, where $p(K_{ij}) \equiv [1 - \exp(-2K_{ij})]$. The presence of different sorts of bonds in model 2 thus leads to different bond probabilities but poses no further problems. If the bond is active, sites $i$ and $j$ belong to the same cluster. The simplest way to simulate this is to draw a random number for each bond and check whether it is smaller than $p(K_{ij})\delta_{s_i,s_j}$. Following this procedure, the speed of the algorithm decreases as the number of interacting neighbors increases. When the couplings are small, a more efficient procedure is possible. As a first step in the SW or LC cluster formation process one obtains, for each type of bond $K_{ij}$, a list of bonds that should be activated if they connect equal spins. To this purpose, one introduces bond variables $b_{ij} = 0$ or 1; the probability that $b_{ij} = 1$ is equal to $p(K_{ij})$. The distribution $P(k) \equiv p(1-p)^{k-1}$, where we write $p$ as an abbreviation for $p(K_{ij})$, expresses the probability that $(k-1)$ subsequent bond variables equal zero, while the $k$-th bond variable is one. Thus one random number $r$ can be transformed into an integer $k$:

$$k = 1 + [\ln(r)/\ln(1-p)],  \qquad (3)$$

where the square brackets denote the integer part. After evaluation of $k$, the next $(k-1)$ entries in the list of bond variables are set to zero, and the $k$-th variable is set to one. By repetition of these steps a complete list of bond variables (for all bonds with strength $K_{ij}$ in the lattice) is obtained. Such lists are generated for each different type of bond. After completion of the lists, the cluster formation is trivial. This procedure was found to improve the speed of the simulation of model 2 considerably. One may still choose between the Swendsen–Wang or largest-cluster method. The latter method was observed to lead to shorter relaxation times and is therefore more efficient. The same principle was applied to Wolff-type simulations of model 2. Random numbers are, as above, transformed into integers $k$. During the cluster formation, $(k-1)$ bonds of the pertinent type are skipped and the spin connected to the $k$-th bond is added to the Wolff cluster if it has the right sign. This leads to a considerably faster Wolff algorithm, in particular because random-number generation is relatively time consuming (see Sec. 3).

In the spin-1 case, transitions between zero and nonzero spin values require special attention. It is not immediately obvious how cluster algorithms could produce these transitions. We follow two different methods for the simulation of the spin-1 model. The first one uses a hybrid algorithm in which Metropolis sweeps alternate with cluster steps. The cluster algorithm acts on the nonzero spins only. Since we do not come close to the tricritical point where the ordered Ising phases meet the spin-zero phase, the regions of zero spins remain limited in size and we do not expect serious critical slowing down due to the equilibration between zero and nonzero spin values.

The second method uses a mapping on a spin-$\frac{1}{2}$ model. We consider a Hamiltonian with two spins $t_i = \pm 1$ and $u_i = \pm 1$ on site $i$ (for all $i$) of the simple cubic lattice:

$$\mathcal{H}_{\mathrm{h}}/k_{\mathrm{B}}T = -M_1 \sum_{\langle ij \rangle}(t_i + u_i)(t_j + u_j) - M_2 \sum_m t_m u_m . \qquad (4)$$

Using the transformation $s_i = (t_i + u_i)/2$ and $v_i = (1 + t_i)(1 - u_i)/4$, the partition function of this model is, up to a constant factor,

$$Z_\mathrm{h} = \sum_{\{s_k\}} \left( \prod_{\langle ij \rangle} \exp[4M_1 s_i s_j] \prod_m \sum_{v_m=0}^{1-|s_m|} \exp[2M_2 s_m^2] \right) , \qquad (5)$$

with $s_i = 0, \pm 1$. Summation over the allowed values of $v_m$ yields a factor 2 if $s_m = 0$. Thus

$$Z_\mathrm{h} = 2^N \sum_{\{s_k\}} \exp\left[ 4M_1 \sum_{\langle ij \rangle} s_i s_j + (2M_2 - \ln 2) \sum_m s_m^2 \right] , \qquad (6)$$

where $N$ denotes the number of spins in the system. This is, apart from the prefactor $2^N$, precisely the partition sum for Eq. (2) for $K_\mathrm{nn} = 4M_1$, $K_\mathrm{3n} = 0$ and $D = \ln 2 - 2M_2$. Equation (4) may thus serve for the application of cluster algorithms to the spin-1 Ising model. The special choice $D = \ln 2$ leads to $M_2 = 0$ so that the spin-$\frac{1}{2}$ Hamiltonian simplifies. We have used three different methods to simulate the spin-1 model: the Metropolis-cluster (MLC) method, the full-cluster (FC) method and the Metropolis-Wolff (MW) method. The MLC method alternates one Metropolis sweep with one largest-cluster inversion, the MW method alternates one Metropolis sweep with 5 or 10 (this choice depends on the system size) Wolff steps. The FC method applies largest-cluster flips to the spin-$\frac{1}{2}$ representation of the model: no Metropolis sweeps are included here.

## 3 Random numbers and consistency tests

Significant systematic errors may be introduced in Monte Carlo simulations by using inadequate random-number generators. It is well known that linear congruential methods based on the truncation of 32-bit integers are unsuitable for long simulations. Even their period of about $10^9$ would be too restrictive. On the other hand, also random-number generators based on binary feedback shift registers may introduce serious errors (see, e.g., Refs. [22, 27–30]). In most cases, the production rule selects two bits from the register and assigns their modulo-2 sum to the new bit. Thus the deviations from randomness are dominated by three-bit correlations. A number of algorithms of this type, using 127-bit shift registers with a period in the order of $10^{38}$, have been rejected on the basis of long tests [23] using Metropolis simulations of the critical Ising model. Recent tests by Ferrenberg *et al.* [31] have shown that such deviations also occur when cluster algorithms are used together with random generators based on a generalized feedback shift register [32].

These findings apply to the Ising model on the square lattice, and use a comparison between simulations and exact results for finite systems. In the case of the three-dimensional Ising model, we have the practical difficulty that no exact results are available for general system sizes. One remaining possibility is a comparison between simulations with different realizations of the random-number generator and/or the spin-updating algorithm. Another possibility to check for systematic deviations is offered by numerical exact results for small systems. However, this test is not sufficient. It has been observed [33] that, in two dimensions, effects due to a random-number generator based on a generalized feedback shift register depend strongly on the system size and may be unobservable in small systems.

Systematic effects in 3-d Ising model simulations are apparent from a comparison between finite-size results for the Binder cumulant [34] reported in Refs. [35] and [19], using the Swendsen–Wang and largest-cluster methods. The random generator used in Ref. [35] was based on a generalized feedback shift register with length 502 [36], and that used in Ref. [19] combined such a generator of length 9689 with a multiplicative rule, by means of bitwise modulo-2 addition. The observed discrepancies may be attributed to the shift-register-based generator with length 502 [36], and become even more prominent in simulations of the Wolff type [36].

It is clear that, for the long simulations implied by the present analysis of the 3-d Ising model, the random-number generators should be selected with great care. A systematic study of biases introduced by shift registers is necessary, in particular the dependence on the system size, shift-register length and the number of correlated bits. From simulations in two dimensions it appeared [37] that the deviations are scalable and become small for large system sizes and register lengths. Thus one may try to suppress systematic effects by using very long feedback shift registers [38]. But here we have chosen for a different method. This choice is based on the observation that the biases decrease when the number of bits in the production rule is increased [33,37]. The bitwise modulo-2 addition of two sequences generated by three-bit production rules usually leads to a sequence in which the dominant correlation is one between 9 bits. Thus, we expected that, using a random generator of this type with sufficiently long registers, the systematic effects would be well below the statistical accuracy, in three dimensions as well as in two. The largest part of the present simulations in three dimensions used the production rules $a_i = a_{i-9218} \oplus a_{i-9689}$ and $b_i = b_{i-97} \oplus b_{i-127}$. These were combined by $r_i = a_i \oplus b_i$, where $a_i$, $b_i$ and $r_i$ are 32-bit integers, and $\oplus$ stands for bitwise modulo-2 addition. Most of the simulations of the nearest-neighbor model reported in Ref. [19] were performed using a random generator which combines a multiplicative sequence with $a_i$. No systematic differences between both types of results were observed, nor were there obvious differences between simulations of the Swendsen–Wang, largest-cluster and Wolff types. Also in the case of models 2 and 3 we checked for the presence of significant differences between the result of the different types of spin-updating algorithms (see Table 3) but none were found. This is consistent with the supposed high quality of the sequence $r_i$.

Therefore, we assume that the sequence $r_i$ is sufficiently uncorrelated, so that the simulation results may serve as a standard, to which data produced by means of other random generators can be compared. Thus, deviations in Wolff simulations of model 1, using 3-bit production rules, were determined and their scaling properties were analyzed [39]. The results are qualitatively the same as in two dimensions, and are completely consistent with the picture that the deviations decrease rapidly with increasing system size and shift-register length. No biases due to correlations of 5 or more bits were observed in Wolff simulations of the three-dimensional Ising model.

As a further test, we have carried out exact numerical calculations of the dimensionless ratio $Q_L(K_{\rm nn})$, the susceptibility $\chi$, the energy-like quantity $S_{\rm nn}$ and the specific-heatlike quantity $c_{\rm nn}$. For more precise definitions we refer to the next sections. These calculations apply to small systems with periodic boundaries: model 1 with sizes $3^3$ and $4^3$, and model 2 with size $3^3$.

The calculations for the $3^3$ systems involve a summation over $2^{27}$ distinct states. In view of the efficiency of the calculation, we divided these states into subsets such that the states in each subset are related by symmetries: spin and spatial inversions, translations and rotations. The number of subsets is 55809, most of which contain 2592 states; this is the number of elements of the symmetry group of the $3^3$ Ising lattice. Once a list of 'independent' states, one per subset, has been made, the calculation becomes rather simple. However, the $4^3$ system is already too time-consuming unless treated carefully. The energy and the specific heat follow simply from the expansion coefficients given by Pearson [40]. For the magnetic quantities $Q$ and $\chi$ we have used a perturbation expansion similar to that described by Saleur and Derrida [41]. Details are given in Ref. [19]. The results are summarized in Table 2. The agreement between the Monte Carlo results and the exact numbers is quite satisfactory and does not suggest any problems with the random-number generator or other defects of the algorithm.

## 4 Test of universality

We have performed extensive simulations of models 1, 2 and 3, using the cluster methods described in Sec. 2. The total simulation time amounts to approximately two years on three workstations. We chose systems with size $L \times L \times L$ and periodic boundaries. The lengths of the runs for the various models and methods are given in Table 3 for each system size.

We sampled and analyzed the dimensionless ratio:

$$Q_L(K_{\rm nn}) = \frac{\langle m^2 \rangle_L^2}{\langle m^4 \rangle_L} \ , \qquad (7)$$

Table 2: Comparison between Monte Carlo and exact results for small system sizes. These data were taken at couplings $K_{\rm nn} = 0.221653$ and $K_{\rm nn} = 0.128006$ for models 1 and 2 respectively.

| $L$ | model # | quantity | MC | exact |
|---|---|---|---|---|
| 3 | 1 | $Q$ | 0.668409 (20) | 0.668427 |
| 3 | 1 | $m^2$ | 0.422978 (19) | 0.422992 |
| 3 | 1 | $S_{\rm nn}$ | 1.434382 (51) | 1.434418 |
| 3 | 1 | $c_{\rm nn}$ | 0.785443 (53) | 0.785413 |
| 4 | 1 | $Q$ | 0.659755 (24) | 0.659779 |
| 4 | 1 | $m^2$ | 0.331228 (16) | 0.331204 |
| 4 | 1 | $S_{\rm nn}$ | 1.293247 (40) | 1.293223 |
| 4 | 1 | $c_{\rm nn}$ | 0.977559 (75) | 0.977575 |
| 3 | 2 | $Q$ | 0.642427 (20) | 0.642415 |
| 3 | 2 | $m^2$ | 0.367413 (17) | 0.367390 |
| 3 | 2 | $S_{\rm nn}$ | 1.134852 (51) | 1.134791 |
| 3 | 2 | $c_{\rm nn}$ | 0.287800 (19) | 0.287772 |

Table 3: Length of Monte Carlo runs in millions of sampled configurations. SW stands for Swendsen–Wang, LC for largest cluster and W for Wolff. For SW and LC, each new configuration corresponds with one cluster decomposition of the lattice. In the Wolff case, 5 (5W) or 10 (10W) Wolff clusters were flipped before a new configuration was used for data taking. For the spin-1 model (model 3) LC, 5W and 10W are preceded by M in order to indicate a Metropolis sweep through the lattice. FC indicates the full-cluster algorithm for the spin-1 model; it flips the largest cluster of a spin-$\frac{1}{2}$ version of the model.

| model | 1 | | | | 2 | | | | 3 | | | |
|---|---|---|---|---|---|---|---|---|---|---|---|---|
| $L$ | SW | LC | 5W | 10W | SW | LC | 5W | 10W | FC | MLC | M5W | M10W |
| 3 | 48 | 52 | 200 | | | 100 | 200 | | 100 | 100 | 300 | |
| 4 | 160 | 40 | 200 | | | 150 | 100 | | 100 | 100 | 300 | |
| 5 | 48 | 52 | 200 | | | 150 | 100 | | 100 | 100 | 400 | |
| 6 | 48 | 52 | 200 | | | 150 | 100 | | 100 | 100 | 400 | |
| 7 | 48 | 52 | 200 | | | 150 | 100 | | 100 | 100 | 400 | |
| 8 | 48 | 52 | 200 | | 10 | 140 | 100 | | 100 | 100 | 400 | |
| 9 | 48 | 52 | 200 | | 10 | 140 | 100 | | 100 | 100 | 300 | |
| 10 | 48 | 52 | 200 | | 10 | 140 | 100 | | 100 | 100 | 300 | |
| 11 | 48 | 52 | 200 | | | 150 | 100 | | 100 | 100 | 300 | |
| 12 | 28 | 72 | 200 | | | 150 | 100 | | 100 | 100 | 200 | |
| 13 | 28 | 72 | 200 | | | 100 | 100 | | | | 250 | |
| 14 | 28 | 72 | 200 | | | 100 | 100 | | | | 200 | |
| 15 | 20 | 30 | | 200 | | 50 | | 100 | | | 200 | |
| 16 | 20 | 30 | | 150 | | 50 | | 100 | | | 200 | |
| 18 | 12 | 38 | | 150 | | 50 | | 50 | | | 120 | |
| 20 | 20 | 10 | | 70 | | 50 | | 50 | | | 120 | |
| 22 | 18 | 12 | | 70 | | 20 | | 80 | | | 120 | |
| 24 | 8 | 12 | | 80 | | 20 | | 80 | | | 150 | |
| 28 | 10 | 10 | | 100 | | 20 | | 80 | | | 50 | 50 |
| 32 | 2 | 18 | | 180 | | 25 | | 75 | | | | 100 |
| 40 | | 10 | | 90 | | 20 | | 80 | | | | 100 |

where $L$ is the finite size of the model and $m$ the magnetization density. We use the renormalization language in order to derive the expected finite-size scaling behavior of $Q_L$. By $f(t, h, u, L^{-1})$ we denote the free-energy density as a function of the temperature and magnetic scaling fields, an irrelevant field and the finite-size field [42,43]. Here, we define the free energy as $F = \ln Z$, so without the normal factor $-1/k_\mathrm{B}T$. Its behavior under renormalization with a scale factor $l$ is

$$f(t, h, u, L^{-1}) = l^{-d} f(l^{y_\mathrm{t}} t, l^{y_\mathrm{h}} h, l^{y_\mathrm{i}} u, l/L) + g(t, h) , \qquad (8)$$

where $y_\mathrm{t}$, $y_\mathrm{h}$ and $y_\mathrm{i}$ are the pertinent renormalization exponents, $d = 3$ is the dimensionality and $g$ is the analytic part of the transformation. By differentiating $k$ times with respect to $h$, and choosing $l = L$ and $h = 0$, one obtains

$$f^{(k)}(t, u, L^{-1}) = L^{k y_\mathrm{h} - d} f^{(k)}(L^{y_\mathrm{t}} t, L^{y_\mathrm{i}} u, 1) + g^{(k)}(t) , \qquad (9)$$

where the dependence on $h$ is no longer needed and therefore suppressed. The expectation values of the second and fourth magnetization moments require differentiations of the free energy with respect to the physical magnetic field $H$:

$$\langle m^2 \rangle = L^{-d} \left( \frac{\partial^2 f}{\partial H^2} \right)_{H=0} \qquad (10)$$

and

$$\langle m^4 \rangle = L^{-3d} \left( \frac{\partial^4 f}{\partial H^4} \right)_{H=0} + 3 L^{-2d} \left( \frac{\partial^2 f}{\partial H^2} \right)_{H=0}^2 . \qquad (11)$$

The Ising up–down symmetry implies that $h$ is an odd function of $H$. Thus the correspondence between the derivatives with respect to $h$ and $H$ is

$$\frac{\partial^2 f}{\partial H^2} = f^{(2)} \left( \frac{\partial h}{\partial H} \right)^2 \qquad (12)$$

and

$$\frac{\partial^4 f}{\partial H^4} = f^{(4)} \left( \frac{\partial h}{\partial H} \right)^4 + 4 f^{(2)} \frac{\partial h}{\partial H} \frac{\partial^3 h}{\partial H^3} , \qquad (13)$$

where, as before, $f^{(k)}$ stands for $\partial^k f / \partial h^k$ and all derivatives with respect to $H$ are evaluated at $H = 0$. In the vicinity of the finite-size limit ($t$ small and $L$ finite), we may Taylor-expand the right-hand side of Eq. (9) in $t$ and $u$. After the appropriate substitutions, the finite-size expansion of $Q_L(K_\mathrm{nn})$ follows as:

$$Q_L(K_\mathrm{nn}) = Q + a_1(K_\mathrm{nn} - K_\mathrm{c}) L^{y_\mathrm{t}} + a_2(K_\mathrm{nn} - K_\mathrm{c})^2 L^{2 y_\mathrm{t}} + a_3(K_\mathrm{nn} - K_\mathrm{c})^3 L^{3 y_\mathrm{t}} + \cdots + b_1 L^{y_\mathrm{i}} + b_2 L^{y_2} + \cdots , \qquad (14)$$

where the $a_i$ and $b_i$ are nonuniversal coefficients and $y_2 = d - 2 y_\mathrm{h}$. The last term is due to the field dependence of the analytic part $g$ in Eq. (9). The nonlinear dependence of $h$ on $H$ leads to even more rapidly decaying contributions (not shown). Terms of the same form, but with different exponents, may be due to other irrelevant fields. Because powers of the geometric factor $\partial h/\partial H$ cancel in the first term, $Q$ is a universal constant (related to the Binder cumulant [34]).

The bulk of the numerical data were taken at couplings $K_\mathrm{nn} = 0.221653, 0.128006$ and $0.393410$ for models 1, 2 and 3 respectively, close to the critical points. The results in terms of $Q_L$ are shown in Table 4. A few points at somewhat different couplings were included in order to estimate the coefficients $a_i$ in Eq. (14). The procedure of the analysis is as follows. We computed $Q_L(K_\mathrm{nn})$ for several values of $L$, $K_\mathrm{nn}$ (near the critical points $K_\mathrm{c}$) for the three models and fitted Eq. (14) to the data. The following parameters were used as input: $y_\mathrm{t} = 1.584$ (4) (from $\varepsilon$-expansion [5]; because the data were taken at couplings so close to the critical points, the results of the fits are practically independent of the precise value); $y_\mathrm{i} = -0.83$ (5) (from series expansions [7]; the fit is rather sensitive to the precise value) and $y_2 = -1.963$ (3) (from renormalization arguments given above and the $\varepsilon$-expansion result [5] for the magnetic exponent; the fit is insensitive to the precise value). The results are summarized in Table 5. It is stressed that the error margins quoted here include the uncertainty due to the possible variations in $y_\mathrm{i}$, $y_\mathrm{t}$ and $y_2$ ($y_\mathrm{h}$). The fits for model 1 indicated that system sizes $L < 7$ should be discarded; they reveal finite-size effects not included in Eq. (14), exceeding the statistical error margins. The fits for models 2 and 3, which exhibit much smaller finite-size effects, include system sizes $L \geq 6$. The fit for model 2 clearly reveals a correction with exponent $y_2 \approx -1.96$. In fact, the large residuals in the absence of such a correction demonstrated its presence. As indicated above, this correction may arise from the analytic part of the transformation, although we cannot exclude contributions due to a second irrelevant exponent. Since there is no obvious reason why this term should be absent in general, we have included it in the fitting procedures for models 1 and 3 as well. Furthermore, we observe that the amplitude $b_1$ of the leading correction to scaling can be suppressed. This amplitude has become quite small in the spin-1 model (model 3) and has even changed sign in model 2. In model 1, the amplitude $b_1$ is relatively large and we have attempted to determine the irrelevant exponent by including it as a parameter in the fit. However, for an acceptable fit it was necessary to include the correction term $b_2 L^{y_2}$. Unfortunately, this frustrated the determination of $y_i$ for model 1: if we fixed $y_2 = -1.963$ the exponent $y_i$ shifted towards $y_2$ and if we included both $y_i$ and $y_2$ as free parameters, they approached the same value.

Table 4: Numerical results for the dimensionless ratio $Q_L = \langle m^2 \rangle_L^2 / \langle m^4 \rangle_L$ for the three Ising models defined in Sec. 2. These data were taken at couplings $K_{\rm nn} = 0.221653$, $0.128006$ and $0.393410$ for models 1, 2 and 3, respectively.

| $L$ | model 1 | model 2 | model 3 |
|---|---|---|---|
| 3 | 0.66839 (2) | 0.64244 (3) | 0.61894 (2) |
| 4 | 0.65976 (2) | 0.63164 (3) | 0.62134 (2) |
| 5 | 0.65373 (2) | 0.62642 (3) | 0.62242 (2) |
| 6 | 0.64919 (2) | 0.62370 (3) | 0.62273 (2) |
| 7 | 0.64579 (3) | 0.62217 (3) | 0.62277 (2) |
| 8 | 0.64318 (3) | 0.62126 (3) | 0.62288 (2) |
| 9 | 0.64107 (3) | 0.62087 (3) | 0.62280 (3) |
| 10 | 0.63943 (3) | 0.62051 (4) | 0.62280 (3) |
| 11 | 0.63803 (3) | 0.62045 (4) | 0.62275 (3) |
| 12 | 0.63688 (3) | 0.62026 (4) | 0.62272 (3) |
| 13 | 0.63591 (4) | 0.62030 (4) | 0.62267 (5) |
| 14 | 0.63514 (4) | 0.62034 (5) | 0.62262 (5) |
| 15 | 0.63441 (3) | 0.62041 (4) | 0.62266 (5) |
| 16 | 0.63376 (4) | 0.62050 (5) | 0.62247 (5) |
| 18 | 0.63270 (4) | 0.62057 (6) | 0.62248 (7) |
| 20 | 0.63187 (6) | 0.62081 (6) | 0.62230 (8) |
| 22 | 0.63117 (6) | 0.62101 (6) | 0.62211 (8) |
| 24 | 0.63052 (6) | 0.62098 (7) | 0.62201 (7) |
| 28 | 0.62958 (6) | 0.62142 (7) | 0.62177 (9) |
| 32 | 0.62879 (5) | 0.62174 (8) | 0.62129 (8) |
| 40 | 0.62761 (8) | 0.62250 (9) | 0.62050 (9) |

Table 5: Results of a data analysis of the three models, including system sizes $L \geq 7$ for model 1 and $L \geq 6$ for models 2 and 3. Besides the ratio $Q$, the critical couplings $K_c$ and the nonuniversal coefficients $a_1$, $a_2$, $b_1$ and $b_2$ are listed.

|  | model 1 | model 2 | model 3 |
|---|---|---|---|
| $Q$ | 0.6232 (8) | 0.6229 (3) | 0.6231 (2) |
| $K_c$ | 0.2216542 (8) | 0.1280034 (4) | 0.3934214 (8) |
| $a_1$ | 0.862 (10) | 1.43 (4) | 0.659 (6) |
| $a_2$ | 0.54 (6) | 1.5 (2) | 0.352 (15) |
| $b_1$ | 0.102 (10) | $-0.043$ (4) | 0.001 (2) |
| $b_2$ | 0.11 (3) | 0.351 (13) | $-0.018$ (9) |

In order to take into account the finite-size effects revealed by the system sizes omitted in the previous fits, we have repeated our data analysis with an additional correction to scaling $b_3 L^{y_3}$ in Eq. (14), where $y_3 = -2y_h$. This term, which is due to the nonlinear dependence of the magnetic scaling field on the physical magnetic field, enabled us to include all system sizes $L \geq 5$ for models 1, 2 and 3 in the analysis. The results, which are presented in Table 6, are consistent with those obtained previously. Again, the error margins quoted include the uncertainty due to the errors in $y_i$, $y_t$ and $y_h$. These data satisfy universality within a margin of less than $10^{-3}$. To our knowledge, this is the most precise verification sofar for 3-d Isinglike models.

Table 6: Results of a data analysis of the three models, where all system sizes $L \geq 5$ were used and a third correction term was included. Besides the ratio $Q$, the nonuniversal coefficients $a_1$, $a_2$, $b_1$, $b_2$ and $b_3$ and the critical couplings $K_c$ are listed.

|       | model 1       | model 2       | model 3        |
|-------|---------------|---------------|----------------|
| $Q$   | 0.6235 (7)    | 0.6231 (4)    | 0.6235 (3)     |
| $K_c$ | 0.2216547 (8) | 0.1280036 (5) | 0.3934224 (10) |
| $a_1$ | 0.862 (9)     | 1.43 (4)      | 0.659 (6)      |
| $a_2$ | 0.54 (6)      | 1.5 (2)       | 0.352 (15)     |
| $b_1$ | 0.098 (9)     | $-0.045$ (6)  | $-0.004$ (3)   |
| $b_2$ | 0.15 (4)      | 0.37 (2)      | 0.02 (2)       |
| $b_3$ | $-4.9$ (8)    | $-1.4$ (8)    | $-2.0$ (7)     |

## 5 Determination of the critical dimensions

This section presents finite-size analyses of the energy, specific heat, spin–spin correlations over half the system size, susceptibility, the temperature derivative of the susceptibility, and the temperature derivative of the ratio $Q_L$. Taking $h = 0$ and $l = L$ in Eq. (8) leads to

$$f(t, u, L^{-1}) = L^{-d} f(L^{y_t} t, L^{y_i} u, 1) + g(t) . \tag{15}$$

Expansion in $t$ and $u$ yields

$$\begin{aligned} f(t, u, L^{-1}) &= L^{-d} \left( f^{(0,0)} + f^{(1,0)} L^{y_t} t + \frac{1}{2} f^{(2,0)} L^{2y_t} t^2 + \cdots + f^{(0,1)} L^{y_i} u + f^{(1,1)} L^{y_t+y_i} tu + \cdots \right) \\ &\quad + g^{(0)} + g^{(1)} t + \frac{1}{2} g^{(2)} t^2 + \cdots , \end{aligned} \tag{16}$$

where $f^{(k,l)}$ stands for $\partial^{k+l} f / \partial^k t \, \partial^l u$. The finite-size scaling behavior of the energy and that of the specific heat follow by differentiation.

## 5.1 The energy

During the simulations, the nearest-neighbor sum $S_{\rm nn} = \sum_{\langle {\rm nn} \rangle} s_i s_j$ was sampled. For model 1, this sum is proportional to the energy; for models 2 and 3 its scaling behavior is similar. Its expectation value is equal to

$$\langle S_{\rm nn} \rangle = \frac{\partial f}{\partial K_{\rm nn}} = \frac{\partial f}{\partial t} \frac{\partial t}{\partial K_{\rm nn}} + \frac{\partial f}{\partial u} \frac{\partial u}{\partial K_{\rm nn}} \ . \tag{17}$$

The finite-size scaling behavior of this quantity thus follows by differentiating Eq. (16) and substitution in Eq. (17):

$$\begin{aligned}\langle S_{\rm nn} \rangle &=& c_0 + c_1(K_{\rm nn} - K_{\rm c}) + \cdots \\ && + L^{y_t - d} \left[ a_0 + a_1(K_{\rm nn} - K_{\rm c}) L^{y_t} + a_2 (K_{\rm nn} - K_{\rm c})^2 L^{2y_t} + \cdots + b_1 L^{y_i} + b_2 L^{y_i - y_t} + \cdots \right] \ , \end{aligned} \tag{18}$$

where the $a_i$, $b_i$ and $c_i$ are unknown coefficients. Analysis of the numerical results for $\langle S_{\rm nn} \rangle$ enables a determination of these coefficients and of $y_t$. The dominant singular term in Eq. (18) is the one with amplitude $a_0$. The $(K_{\rm nn} - K_{\rm c})$-dependent term with amplitude $c_1$ is dominated by the term with coefficient $a_1$ and has therefore been omitted from the scaling formula. Since the bulk of the data were taken very close to the critical points, only linear and quadratic terms in $(K_{\rm nn} - K_{\rm c})$ were included. Without the correction term with coefficient $b_2$, we had to exclude system sizes $L < 8$ in the analysis of model 1, in order to obtain an acceptable residual. The resulting estimate for $y_t$ is: 1.586 (6). Inclusion of the second irrelevant term enabled us to include all system sizes $L \geq 5$. For consistency, we have included this term in the data analyses for models 2 and 3 as well. Table 7 summarizes the results obtained from fits according to Eq. (18), at the critical points listed in Table 6, for system sizes $L \geq 5$. Since the singular behavior of $\langle S_{\rm nn} \rangle$ is rather weak, the results $y_t \approx 1.59$ for each of the three models are relatively inaccurate but consistent with the existing literature. The uncertainty due to the errors in $K_{\rm c}$ and $y_i$ has been included in the error margins.

Table 7: Results of a data analysis of the nearest-neighbor sum $\langle S_{\rm nn} \rangle$ for the three models.

|       | model 1     | model 2     | model 3     |
|-------|-------------|-------------|-------------|
| $y_t$ | 1.599 (8)   | 1.589 (9)   | 1.591 (7)   |
| $c_0$ | 0.99051 (8) | 0.66298 (9) | 0.59451 (6) |
| $a_0$ | 2.14 (6)    | 2.20 (7)    | 1.73 (4)    |
| $b_1$ | 0.14 (15)   | 0.16 (17)   | 0.04 (12)   |
| $b_2$ | $-2.0$ (4)  | $-0.6$ (4)  | $-0.9$ (3)  |

## 5.2 The specific heat

The fluctuations in $S_{\rm nn}$ are related to the specific-heatlike quantity

$$c_{\rm nn} = K_{\rm nn}^2 \frac{\partial^2 f}{\partial K_{\rm nn}^2} = K_{\rm nn}^2 \left[ \langle S_{\rm nn}^2 \rangle - \langle S_{\rm nn} \rangle^2 \right] \ . \tag{19}$$

We consider $f$ as a function of the scaling fields $t$ and $u$:

$$c_{\rm nn} = K_{\rm nn}^2 \left[ \frac{\partial f}{\partial t} \frac{\partial^2 t}{\partial K_{\rm nn}^2} + \frac{\partial f}{\partial u} \frac{\partial^2 u}{\partial K_{\rm nn}^2} + \frac{\partial^2 f}{\partial t^2} \left( \frac{\partial t}{\partial K_{\rm nn}} \right)^2 + 2 \frac{\partial^2 f}{\partial t \partial u} \frac{\partial t}{\partial K_{\rm nn}} \frac{\partial u}{\partial K_{\rm nn}} + \frac{\partial^2 f}{\partial u^2} \left( \frac{\partial u}{\partial K_{\rm nn}} \right)^2 \right] \ . \tag{20}$$

Taking the appropriate derivatives in Eq. (16) and collecting the leading analytic and singular terms leads to

$$\begin{aligned} c_{\rm nn} &=& p_0 + p_1 (K_{\rm nn} - K_{\rm c}) + \cdots \\ && + L^{2y_t - d} \left[ q_0 + q_1 (K_{\rm nn} - K_{\rm c}) L^{y_t} + q_2 (K_{\rm nn} - K_{\rm c})^2 L^{2y_t} + \cdots + r_1 L^{y_i} + \cdots \right] \\ && + L^{y_t - d} \left[ s_0 + s_1 (K_{\rm nn} - K_{\rm c}) L^{y_t} + \cdots \right] \ . \end{aligned} \tag{21}$$

The numerical results for $c_{\rm nn}$ of models 1–3 were subjected to a fit of this form with $y_{\rm i} = -0.83$ and the $K_{\rm c}$ values in Table 6 as input parameters. The terms with amplitudes $p_1$ and $s_1$ are dominated by that with amplitude $q_1$ and were omitted from the fit formula, as well as quadratic terms in $(K_{\rm nn} - K_{\rm c})$. System sizes $L < 6$ display finite-size corrections not included in Eq. (21) and were discarded. The main results of these fits are shown in Table 8, where the error margins include the uncertainties in $K_{\rm c}$ and $y_{\rm i}$. Also in the present case we find consistent, but inaccurate values of $y_{\rm t}$. This may be related to the fact that the leading power of $L$ in Eq. (21) is close to zero, so that this term, which has the coefficient $q_0$, interferes with the term with coefficient $p_0$.

Table 8: Results of a data analysis of the specific-heatlike quantity $c_{\rm nn}$ obtained from the fluctuations of the nearest-neighbor sum $S_{\rm nn}$ for each of the three models.

|  | model 1 | model 2 | model 3 |
|---|---|---|---|
| $y_{\rm t}$ | 1.60 (2) | 1.579 (15) | 1.59 (2) |
| $p_0$ | −0.8 (7) | −0.6 (3) | −3 (2) |
| $q_0$ | 1.5 (5) | 0.8 (3) | 3.5 (14) |
| $q_1$ | 2.2 (2) | 1.39 (12) | 3.7 (4) |
| $r_1$ | −0.4 (4) | 0.11 (14) | 0.0 (9) |
| $s_0$ | −0.4 (3) | −0.24 (13) | −1.0 (7) |

## 5.3 The spin–spin correlation function

In our simulations, we have sampled the spin–spin correlation function $g(\boldsymbol{r})$,

$$g(\boldsymbol{r}) = \langle s(\boldsymbol{0}) s(\boldsymbol{r}) \rangle \;, \tag{22}$$

over half the system size ($r = L/2$), for even system sizes. This quantity can be derived from the free energy $F$ by differentiating with respect to two physical magnetic fields $H_0$ and $H_r$, which couple to the spins at positions $\boldsymbol{0}$ and $\boldsymbol{r}$, respectively. We consider the two fields as independent and find

$$g(\boldsymbol{r}) = \left( \frac{\partial^2 F}{\partial H_0 \partial H_r} \right)_{H_0 = H_r = 0} = \left( \frac{\partial^2 F}{\partial h_0 \partial h_r} \frac{\partial h_0}{\partial H_0} \frac{\partial h_r}{\partial H_r} \right)_{H_0 = H_r = 0} \;, \tag{23}$$

where $h$ denotes the leading magnetic scaling field and the derivatives with respect to this field are evaluated at $h_0 = h_r = 0$. Using Eq. (9) one obtains upon expansion in $t$ and $u$ the scaling behavior of the correlation function,

$$g = L^{2y_{\rm h} - 2d} \left[ a_0 + a_1 (K_{\rm nn} - K_{\rm c}) L^{y_{\rm t}} + a_2 (K_{\rm nn} - K_{\rm c})^2 L^{2y_{\rm t}} + \cdots + b_1 L^{y_{\rm i}} + \cdots \right] \;, \tag{24}$$

where the coefficients $a_i$ and $b_i$ are different from those in Eq. (18).

We have fitted the terms shown in (24) to our data. The large residuals for all three models strongly suggested the presence of an additional correction to scaling $b_2 L^{y'}$. A problem for the determination of $y'$ is the presence of the leading correction term $b_1 L^{y_{\rm i}}$. Only in the spin-1 model (model 3), where the amplitude $b_1$ is small and the term thus may be omitted, a reasonable determination was possible, yielding $y' = -2.1$ (1). This could be a second temperature-like irrelevant exponent, although we haven't observed it in the analysis of the ratio $Q$ or the energy-like quantity $S_{\rm nn}$. In $Q$, it may have been masked by the term $b_2 L^{y_2}$, but this is less likely for $S_{\rm nn}$, where the exponent of the second correction term is approximately equal to $-2.4$. On the other hand, the contribution $b_2 L^{y'}$ could, in principle, be due to a second relevant magnetic exponent $\tilde{y}_{\rm h}$. Taking into account the dependence of $F$ on an additional magnetic scaling field $\tilde{h}$ yields

$$g(\boldsymbol{r}) = \frac{\partial^2 F}{\partial h_0 \partial h_r} \left( \frac{\partial h}{\partial H} \right)^2 + \left( \frac{\partial^2 F}{\partial h_0 \partial \tilde{h}_r} + \frac{\partial^2 F}{\partial \tilde{h}_0 \partial h_r} \right) \frac{\partial h}{\partial H} \frac{\partial \tilde{h}}{\partial H} + \frac{\partial^2 F}{\partial \tilde{h}_0 \partial \tilde{h}_r} \left( \frac{\partial \tilde{h}}{\partial H} \right)^2 \;. \tag{25}$$

This results in extra terms proportional to $L^{y_h+\tilde{y}_h-2d}$ and $L^{2\tilde{y}_h-2d}$ in the scaling formula for $g$, corresponding to correction terms $L^{\tilde{y}_h-y_h}$ and $L^{2\tilde{y}_h-2y_h}$ in Eq. (24). Remarkably, the second magnetic exponent $\tilde{y}_h = 0.42$ [14, 18] has just the right value. However, its identification in terms of a redundant operator [14, 18] would exclude its contribution to thermodynamic quantities. Table 9 shows the main results of an analysis for system sizes $L \geq 8$, where we have included three correction terms, $b_1 L^{y_i}$, $b_2 L^{y'}$ and $b_3 L^{2y'}$, with the exponents $y_i$ and $y'$ fixed at $-0.83$ and $-2.1$, respectively. We have not included the term proportional to $L^{2y_i}$ because $b_1$ is already quite small. The errors quoted in the table include the uncertainties in $K_c$, $y_t$, $y_i$ and $y'$. The estimates of $y_h$ for each of the three models are consistent and in agreement with the existing literature.

Table 9: Results of a data analysis of the spin–spin correlation function $g$ for the three models.

|       | model 1    | model 2    | model 3    |
|-------|------------|------------|------------|
| $y_h$ | 2.480 (2)  | 2.482 (3)  | 2.482 (3)  |
| $a_0$ | 0.77 (2)   | 0.547 (15) | 0.453 (14) |
| $a_1$ | 2.45 (4)   | 3.07 (6)   | 1.18 (2)   |
| $a_2$ | 3.44 (14)  | 7.2 (3)    | 1.30 (3)   |
| $b_1$ | $-0.22$ (10) | 0.08 (8) | 0.01 (7)   |

## 5.4 The magnetic susceptibility

The magnetic susceptibility $\chi$ can be calculated from the average square magnetization, which is sampled in the Monte Carlo simulations,

$$\chi = L^d \langle m^2 \rangle . \tag{26}$$

Using Eqs. (9), (10) and (12), we find for the finite-size scaling behavior:

$$\chi = g^{(2)}(t) + L^{2y_h-d} f^{(2)}(L^{y_t}t, L^{y_i}u, 1) , \tag{27}$$

which yields, upon expansion in $t$ and $u$,

$$\chi = c_0 + c_1(K_{nn} - K_c) + \cdots + L^{2y_h-d}\left[a_0 + a_1(K_{nn} - K_c)L^{y_t} + a_2(K_{nn} - K_c)L^{2y_t} + b_1 L^{y_i} + \cdots\right] , \tag{28}$$

where the $a_i$, $b_i$ and $c_i$ are nonuniversal coefficients. In Table 10, we present the results of fits of the susceptibility for the models 1–3 at the critical points listed in Table 6. For model 1, system sizes $L \geq 8$ were included in the analysis and for models 2 and 3, which exhibit smaller corrections to scaling, all system sizes $L \geq 6$ were used. The coefficient $c_1$ in Eq. (28) was set to zero in all analyses, because the term containing it is much smaller than the $(K_{nn} - K_c)$-dependent term with amplitude $a_1$. The errors include the margins due to the uncertainties in $K_c$, $y_i$ and $y_t$. The ratio between the coefficients $a_0$ for the three models is in excellent agreement with the ratio between the coefficients $a_0$ in Table 9. The same holds for the coefficients $a_1$, $a_2$ and $b_1$.

On the other hand, one might derive the scaling formula for $\chi$ from that of the spin–spin correlation function $g$, because $\chi$ is equal to the spatial integral of $g$,

$$\chi = \int g(r) r^{d-1} \, dr . \tag{29}$$

Since the integral in Eq. (29) preserves the form of the corrections to scaling in $g$, we expect the same type of corrections in the correlation function and the susceptibility. Only the terms proportional to $c_0, c_1, \ldots$ in (28), which arise from the analytical part of the free energy, are absent in Eq. (24). These contributions come from the small-$r$ cutoff in Eq. (29). Thus, we have included in the scaling formula the additional corrections that we observed in the analysis of the correlation function. As the term proportional to $L^{y'}$ interferes with the constant contribution $c_0$, we have only included the correction $b_2 L^{2y'}$. This allowed us to include system sizes $L \geq 5$ for all three models. The results, which are presented in Table 11, are consistent with those obtained in the previous analysis. Now, the errors also include the margins due to the uncertainty in $y'$. Just as in the analysis of the correlation function, we find consistent results for $y_h$, which are in agreement with the literature. However, the values for $y_h$ are more accurate than those obtained in the previous subsection and our resulting estimate for the magnetic renormalization exponent is $y_h = 2.4815\,(15)$. The error margin amounts to two standard deviations, in order to take into account any arbitrariness in the fit formula.

Table 10: Results of a data analysis of the susceptibility $\chi$ for the three models.

|       | model 1     | model 2     | model 3     |
|-------|-------------|-------------|-------------|
| $y_h$ | 2.4812 (11) | 2.4817 (10) | 2.4826 (9)  |
| $c_0$ | $-0.6$ (2)  | $-0.20$ (7) | $-0.50$ (6) |
| $a_0$ | 1.559 (16)  | 1.126 (9)   | 0.926 (7)   |
| $a_1$ | 4.88 (6)    | 6.16 (8)    | 2.36 (3)    |
| $a_2$ | 6.9 (4)     | 14.4 (7)    | 2.62 (7)    |
| $b_1$ | $-0.37$ (5) | 0.14 (3)    | $-0.05$ (2) |

Table 11: Results of a data analysis of the susceptibility $\chi$ for the three models, where all system sizes $L \geq 5$ were employed and an additional correction to scaling was included in the scaling formula.

|       | model 1     | model 2     | model 3     |
|-------|-------------|-------------|-------------|
| $y_h$ | 2.4813 (11) | 2.4810 (14) | 2.4817 (13) |
| $c_0$ | $-0.5$ (2)  | 0.0 (2)     | $-0.33$ (13)|
| $a_0$ | 1.558 (15)  | 1.134 (13)  | 0.934 (10)  |
| $a_1$ | 4.87 (6)    | 6.18 (8)    | 2.37 (3)    |
| $a_2$ | 6.9 (3)     | 14.5 (7)    | 2.64 (7)    |
| $b_1$ | $-0.37$ (5) | 0.10 (6)    | 0.01 (13)   |
| $b_2$ | $-5$ (2)    | $-2.9$ (16) | $-2.6$ (13) |

## 5.5 The temperature derivative of $\chi$

In the simulations, we have also sampled the correlation between $m^2$ and $S_{\rm nn}$. This allows us to calculate the temperature derivative of the susceptibility,

$$\frac{\partial \chi}{\partial K_{\rm nn}} = L^d \left( \langle m^2 S_{\rm nn} \rangle - \langle m^2 \rangle \langle S_{\rm nn} \rangle \right) \ . \tag{30}$$

The scaling behavior of this quantity can be derived directly from that of the susceptibility, Eq. (28),

$$\frac{\partial \chi}{\partial K_{\rm nn}} = c_1 + \cdots + L^{2y_h + y_t - d} \left[ a_1 + 2a_2 (K_{\rm nn} - K_{\rm c}) L^{y_t} + 3a_3 (K_{\rm nn} - K_{\rm c})^2 L^{2y_t} + \cdots + \tilde{b}_1 L^{y_i} + \cdots \right] \ . \tag{31}$$

The term with amplitude $\tilde{b}_1$ comes from a term proportional to $(K_{\rm nn} - K_{\rm c}) L^{y_t + y_i}$, included in the ellipsis in Eq. (28). Just as in the analysis of the spin–spin correlation function, the residuals for all three models indicated the presence of an additional correction to scaling $\tilde{b}_2 L^{y'}$, which indeed follows from the discussion in the previous subsection. Table 12 shows the results of an analysis at the critical points listed in Table 6, where this additional correction was included. All system sizes $L \geq 6$ were used. The exponents $y_i$ and $y'$ were kept fixed at $-0.83$ and $-2.1$, respectively. The error margins include the uncertainties in $K_{\rm c}$, $y_i$ and $y'$. The fit yields values for $(2y_h + y_t)$ and the results for $y_t$ have been obtained by fixing $y_h$ at the best estimate from the previous subsection. This implies an additional error margin of 0.003 for $y_t$. For models 1 and 3, there is a reasonable agreement between the amplitudes $a_1$ and $a_2$ as shown in Table 11 and those in Table 12. The differences are explained from the approximations in the scaling formulae. For model 2, no agreement is expected, because an additional term arises in the temperature derivative of the susceptibility due to the temperature dependence of $K_{3n}$ (the ratio between $K_{nn}$ and $K_{3n}$ is fixed).

Table 12: Results of a data analysis of the temperature derivative of the susceptibility $\partial \chi / \partial K_{nn}$ for the three models, where all system sizes $L \geq 6$ were employed and an additional correction to scaling was included in the scaling formula.

|  | model 1 | model 2 | model 3 |
| --- | --- | --- | --- |
| $y_t$ | 1.585 (3) | 1.584 (4) | 1.587 (4) |
| $c_1$ | 31 (11) | 1 (9) | 12 (7) |
| $a_1$ | 5.11 (9) | 3.73 (6) | 2.39 (5) |
| $a_2$ | 6.46 (10) | 8.04 (9) | 2.53 (3) |
| $a_3$ | $-2.4$ (4) | $-4.4$ (5) | $-0.53$ (5) |
| $\tilde{b}_1$ | $-2.6$ (3) | 0.1 (3) | $-0.1$ (2) |
| $\tilde{b}_2$ | $-13$ (3) | $-6$ (2) | $-5.5$ (16) |

## 5.6 The temperature derivative of $Q$

Another quantity of interest correlates the magnetization distribution with the nearest-neighbor sum $S_{nn}$:

$$\frac{\partial Q}{\partial K_{nn}} = Q \left( 2 \frac{\langle m^2 S_{nn}\rangle - \langle m^2\rangle\langle S_{nn}\rangle}{\langle m^2\rangle} - \frac{\langle m^4 S_{nn}\rangle - \langle m^4\rangle\langle S_{nn}\rangle}{\langle m^4\rangle} \right) \; . \tag{32}$$

The determination of $m$ and $S_{nn}$ during the simulations enables the sampling of this quantity with very little additional effort. Returning to Eq. (14) and noting that the ellipses include terms proportional to $(K_{nn} - K_c)L^{y_t+y_i}$ and to $(K_{nn} - K_c)L^{y_2}$, we obtain the finite-size scaling behavior

$$\frac{\partial Q}{\partial K_{nn}} = L^{y_t} \left[ u_0 + u_1(K_{nn} - K_c)L^{y_t} + u_2(K_{nn} - K_c)^2 L^{2y_t} + \cdots + vL^{y_i} + wL^{y_2-y_t} + \cdots \right] \; . \tag{33}$$

The numerical data for the three models were subjected to a fit on the basis of Eq. (33), where we have included system sizes $L \geq 7$ for model 1 and $L \geq 5$ for models 2 and 3. In this case the leading power of $L$ stands well apart from the less singular terms and the results for $y_t$ (Table 13; uncertainties in $K_c$, $y_i$ and $y_h$ are included in all error margins) appear to be more accurate than those in the preceding subsections. The results suggest that the correction due to the leading irrelevant field is very small. Therefore we have repeated our analysis with $v$ fixed to zero. We expect this to work especially well for models 2 and 3, where the irrelevant field is notably smaller than in the first model. Indeed, we have obtained accurate and consistent results for the models 2 and 3, as shown in Table 14. These results, together with those presented in Table 13, lead us to our final result, $y_t = 1.587$ (2). Just as in the final result for $y_h$ in Sec. 5.4, we quote here an error margin of two standard deviations.

## 6 Simultaneous fits for the three models

Considering the results in the preceding sections, it is reasonable to assume now that universality is *exactly* satisfied for the three models under investigation. Thus we made a fit of the combined data for the ratio $Q$, allowing only single values of $Q$ and $y_i$ for the three models. The other parameters $a_1$, $a_2$, $K_c$, $b_1$ and

Table 13: Results of a data analysis of the derivative of the quantity $Q$ with respect to the nearest-neighbor coupling $K_{\rm nn}$.

|       | model 1     | model 2     | model 3      |
|-------|-------------|-------------|--------------|
| $y_{\rm t}$ | 1.589 (2)   | 1.587 (2)   | 1.5878 (14)  |
| $u_0$ | 1.341 (14)  | 1.351 (9)   | 1.057 (6)    |
| $u_1$ | $-0.21$ (3) | 0.00 (5)    | $-0.012$ (9) |
| $u_2$ | $-10.7$ (6) | $-28.3$ (11)| $-4.80$ (12) |
| $v$   | $-0.01$ (6) | 0.00 (3)    | 0.02 (2)     |
| $w$   | $-0.5$ (2)  | 0.46 (8)    | $-0.13$ (5)  |

Table 14: Results of a data analysis of the derivative of the quantity $Q$ with respect to the nearest-neighbor coupling $K_{\rm nn}$ for models 2 and 3, where the leading correction to scaling has been omitted.

|       | model 2      | model 3       |
|-------|--------------|---------------|
| $y_{\rm t}$ | 1.5868 (3)   | 1.5867 (2)    |
| $u_0$ | 1.3512 (11)  | 1.0623 (7)    |
| $u_1$ | 0.00 (5)     | $-0.013$ (9)  |
| $u_2$ | $-28.3$ (11) | $-4.81$ (12)  |
| $w$   | 0.457 (15)   | $-0.091$ (9)  |

$b_2$ (see Eq. (14)) are nonuniversal and occur in triplicate. Now, system sizes $L < 8$ had to be discarded, except when an additional correction to scaling proportional to $L^{-2y_{\rm h}}$ was added to the scaling formula. In the latter case, all system sizes $L \geq 5$ could be included. Some of the results are summarized in Tables 15 and 16, respectively, where the error margins include the uncertainty introduced by the error in $y_{\rm t}$ and $y_{\rm h}$.

Table 15: Results of a data analysis assuming universality of $Q_L = \langle m^2 \rangle_L^2 / \langle m^4 \rangle_L$ for the three investigated Ising models. System sizes $L \geq 8$ were included in the fit. The table lists nonuniversal parameters: the critical points and the amplitudes of the *two* correction terms. Furthermore, this analysis yielded the universal parameters $Q = 0.6232$ (2) and $y_{\rm i} = -0.78$ (3).

|       | model 1         | model 2         | model 3         |
|-------|-----------------|-----------------|-----------------|
| $K_{\rm c}$ | 0.2216550 (6)   | 0.1280037 (4)   | 0.3934217 (8)   |
| $b_1$ | 0.086 (8)       | $-0.040$ (5)    | $-0.001$ (2)    |
| $b_2$ | 0.18 (3)        | 0.34 (2)        | 0.000 (14)      |

Let us now compare the results of the various fits. In the first place, we see that the results in Tables 15 and 16 are consistent, just as was the case for Tables 5 and 6 in Sec. 4. Also the values for the universal quantity $Q$, 0.6232 (2) and 0.6233 (2), respectively, agree. Secondly, the simultaneous fit with only the first two corrections to scaling (Table 15) yields results that are consistent with those presented in Table 5. Only the amplitude $b_2$ and the critical coupling $K_{\rm c}$ for model 1 appear to be somewhat too low in Table 5, as we already had seen from the second fit in Sec. 4. Finally, when we compare the results in Tables 6 and 16, i.e., including a third correction to scaling, as well as the corresponding $Q$ values, we see a very good agreement. These comparisons, in addition to the fact that the term $b_3 L^{y_3}$ allowed us to include all system sizes $L \geq 5$, lead us to the conclusion that the fits presented in Table 16 can be considered as the most accurate results. In addition to the nonuniversal constants given in the table and the universal amplitude ratio $Q$, this analysis yielded the (universal) irrelevant exponent $y_{\rm i} = -0.82$ (3). This value is in very good agreement with that obtained by Nickel and Rehr [7]. Although there is one more unknown ($y_i$), the results for $Q$ and $K_c$ obtained in this section are more accurate than those of the three separate fits. One of the reasons is that the fit for model 3 is insensitive to the value of $y_i$, so that, e.g., $Q$ is determined accurately.

Table 16: Results of a data analysis assuming universality of $Q_L = \langle m^2 \rangle_L^2 / \langle m^4 \rangle_L$ for the three investigated Ising models. System sizes $L \geq 5$ were included in the fit. The table lists nonuniversal parameters: the critical points and the amplitudes of the *three* correction terms. Furthermore, this analysis yielded the universal parameters $Q = 0.6233\ (2)$ and $y_i = -0.82\ (3)$.

|       | model 1       | model 2       | model 3       |
|-------|---------------|---------------|---------------|
| $K_c$ | 0.2216546 (5) | 0.1280039 (4) | 0.3934220 (7) |
| $b_1$ | 0.096 (7)     | $-0.046$ (5)  | $-0.002$ (2)  |
| $b_2$ | 0.15 (3)      | 0.38 (2)      | 0.007 (12)    |
| $b_3$ | $-4.9$ (8)    | $-1.7$ (7)    | $-1.7$ (5)    |

# 7 Discussion and conclusion

Let us summarize our final results for the renormalization exponents: $y_t = 1.587\ (2)$, $y_h = 2.4815\ (15)$, $y_i = -0.82\ (6)$. To allow for any residual dependencies on the choice of the fitting formulae, we list error margins of two standard deviations. In Table 17 we compare our results with some recent estimates obtained by various methods. Our result for the temperature exponent is lower than that of Baillie *et al.*, obtained by the Monte Carlo renormalization method. This could be explained by a violation of hyperscaling. However, the accurate agreement between our result and that of coupling-constant expansion [2], $\varepsilon$-expansion [5], series expansions [7,8] and the coherent-anomaly method [10] makes this explanation less likely. We notice that our result for $y_t$ is markedly higher than the recent series-expansion result of Guttman and Enting [9]. The result for the magnetic exponent is also in good agreement with most other estimates, although the result of Baillie *et al.* lies significantly higher than the majority of the results. Also the result of Le Guillou and Zinn-Justin obtained by coupling-constant expansion [2] seems somewhat too high. The results for the leading irrelevant exponent are not very accurate, but consistent. We notice that the fractions $\frac{73}{46}$ and $\frac{67}{27}$ are good approximations for $y_t$ and $y_h$, respectively. For easy reference, Table 18 summarizes the exponents $\alpha$, $\beta$, $\gamma$, $\delta$, $\eta$, $\nu$ and $\Delta$ as calculated from our results for $y_t$, $y_h$ and $y_i$, on the assumption that the hypotheses of scaling and hyperscaling are valid.

Table 17: Some recent results for the renormalization exponents.

|                              | Year | $y_t$     | $y_h$       | $y_i$           |
|------------------------------|------|-----------|-------------|-----------------|
| Present work                 | 1995 | 1.587 (2) | 2.4815 (15) | $-0.82$ (6)     |
| Kolesik and Suzuki [10]      | 1995 | 1.586 (4) | 2.482 (4)   |                 |
| Guttman and Enting [9]       | 1994 | 1.580 (3) |             |                 |
| Landau [20]                  | 1994 | 1.590 (2) | 2.482 (7)   |                 |
| Baillie *et al.* [18]        | 1992 | 1.602 (5) | 2.4870 (15) | $-0.8$ to $-0.85$ |
| Nickel [8]                   | 1991 | 1.587     | 2.4823      | $-0.84$         |
| Nickel and Rehr [7]          | 1990 | 1.587 (4) | 2.4821 (4)  | $-0.83$ (5)     |
| Le Guillou and Zinn-Justin [5] | 1987 | 1.584 (4) | 2.4813 (13) |                 |
| Le Guillou and Zinn-Justin [2] | 1980 | 1.587 (4) | 2.485 (2)   | $-0.79$ (3)     |

Furthermore, we can calculate the Binder cumulant $U$ from our estimate for $Q$, using the relation $U = 3 - 1/Q$, which yields $U = 1.3956\ (10)$. Only few accurate results are available for this quantity (see, e.g.,

Table 18: The standard critical exponents as well as Wegner's correction-to-scaling exponent $\Delta$ as calculated from our best estimates for $y_t$, $y_h$ and $y_i$.

| Exponent | Expressed in RG exp. | Value |
|---|---|---|
| $\alpha$ | $2 - d/y_t$ | 0.110 (2) |
| $\beta$ | $(d - y_h)/y_t$ | 0.3267 (10) |
| $\gamma$ | $(2y_h - d)/y_t$ | 1.237 (2) |
| $\delta$ | $y_h/(d - y_h)$ | 4.786 (14) |
| $\eta$ | $2 - 2y_h + d$ | 0.037 (3) |
| $\nu$ | $1/y_t$ | 0.6301 (8) |
| $\Delta$ | $-y_i/y_t$ | 0.52 (4) |

Ref. [44] for a review) and one of the most accurate estimates up till now is $U = 1.403$ (7) [45]. Our result is in agreement with this and other estimates, but its accuracy is markedly higher. We have not sampled the characteristic length $\xi_L$ defined by Baker and Kawashima [46] in our simulations. Thus, we have no result for the renormalized coupling constant $g^*$, which differs from the Binder cumulant by a factor $(L/\xi_L)^d$ [46].

Table 19 presents a comparison of recent results for $K_c$ of the spin-$\frac{1}{2}$ nearest-neighbor Ising model. Again, it should be noted that the error margin of the result obtained in the present work amounts to two standard deviations. It can be seen that the amplitude ratio $Q$, which is used in this work, provides a good means of obtaining an accurate estimate for the critical coupling. We conclude that the conjecture of Rosengren [47] is not correct. The result of Ferrenberg and Landau deviates by 1.8 combined standard errors, but the newest estimate of Landau differs by only 1.2 standard deviations from the result presented here. The difference with Ref. [19] is 1.6 standard errors and is partly due to statistical errors (the data used in this work include those of Ref. [19] but are much more accurate and include $L = 40$ data), and partly because a term with exponent $y_2$ was not included in the scaling formula for the ratio $Q$. Finally, we want to stress the importance of the spin-1 model. Since the corrections to scaling are small in this model, it is very suitable for the determination of universal quantities.

Table 19: Summary of recent results for the critical point of the spin-$\frac{1}{2}$ Ising model with nearest-neighbor couplings.

| Reference | Year | Value |
|---|---|---|
| Present work | 1995 | 0.2216546 (10) |
| Landau [20] | 1994 | 0.2216576 (22) |
| Blöte and Kamieniarz [19] | 1993 | 0.221648 (4) |
| Baillie et al. [18] | 1992 | 0.221652 (3) |
| Livet [35] | 1991 | 0.2216544 (10) |
| Ferrenberg and Landau [17] | 1991 | 0.2216595 (26) |
| Ito and Suzuki [16] | 1991 | 0.221657 (3) |
| Blöte et al. [13] | 1989 | 0.221652 (5) |
| Rosengren (conjecture) [47] | 1986 | 0.2216586 (0) |

## Acknowledgements

We thank A. Compagner for contributing his knowledge of shift-register-based random-number generators. We are much indebted to M.E. Fisher, M.P. Nightingale, L.N. Shchur and A.L. Talapov for valuable discussions, and to F. Livet for supplying the details of the random-number generator used in Ref. [35]. This work is part of the research program of the "Stichting voor Fundamenteel onderzoek der Materie (FOM)" which is financially supported by the "Nederlandse Organisatie voor Wetenschappelijk Onderzoek (NWO)".

# A  Experimental results

In this appendix, we have collected a number of experimental results for the various critical exponents of phase transitions which have been compared to results for the 3-d Ising universality class. Only results published after 1980 are included. For binary mixtures, older results can be found in, e.g., Ref. [48]. The substances are grouped into four different subsets: unary systems, mixtures, magnetic systems and micellar systems (microemulsions). In general, the results agree very well with the theoretical values, but there are several remarkable discrepancies. Here, we only mention those measurements that differ by more than two standard deviations from our results. Results without error estimates are not taken into account.

The value for $\gamma$ found in Ref. [69] lies much below the theoretical estimate. In Ref. [114] the results for both $\gamma$ and $\nu$ do not appear to be Isinglike, as the authors have already noticed. The results for $SF_6$ presented in Ref. [62] are included, because the authors find mean-field values for the critical exponents, whereas the other results for the same substance appear to fall in the Ising universality class. For $\alpha$, a range of values is found in Ref. [73], the lower end of which coincides with the theoretical value. In Ref. [104], two values of $\alpha$ are presented for $MnF_2$, $2.4\sigma$ above and $3.5\sigma$ below the theoretical value, respectively. Several very accurate results for the critical exponent $\beta$ are presented in Refs. [56, 57, 65], which all lie much above the result in Table 18. On the other hand, the accurate results in Refs. [94, 109] lie much lower than the theoretical prediction. The differences in Refs. [51, 83, 85] are less severe, but all results in these references lie between 2 and 3 combined standard errors above the theoretical value. Also the results for $\beta$ presented in Ref. [95] deviate by many standard errors from the theoretical value, although the authors of this reference state that they consider the differences as not significant. The authors of Ref. [72] found a good fit of their data to a value of $\beta = 0.3337$ (5), but conclude that this value is probably too high.

For $\nu$, very low values have been found in Refs. [84, 93], whereas the value in Ref. [80] lies 2.6 standard errors above our result. In Ref. [118], the result $\eta = 0.016$ (5) is given, which deviates by about four combined standard deviations from the best theoretical values. Nevertheless, the authors of Ref. [118] consider it to be in good agreement.

In general, it is difficult to assess the source of the discrepancies noticed here, although there certainly are cases where crossover phenomena and corrections to scaling were not taken into account in the data analysis.

Table 20: The various substances are grouped as follows: 'u' stands for unary systems, 'mx' for mixtures, 'mg' for magnetic systems and 'mi' for micellar systems (microemulsions). $\alpha$, $\beta$, $\gamma$, $\delta$ and $\nu$ denote the standard critical exponents; $\Delta$ is Wegner's correction-to-scaling exponent; the critical exponents which carry a tilde refer to Fisher's 'renormalized' critical exponents [49]. Used abbreviations: AOT = di-2-ethylhexylsulfosuccinate; BHDC = benzyldimethyl-$n$-hexadecylammoniumchloride; $C_iE_j$ = $CH_3 \cdot (CH_2)_{i-1} \cdot (O \cdot CH_2 \cdot CH_2)_j \cdot OH$; HFC-32 = difluoromethane; HFC-125 = pentafluoroethene; R114 = $CClF_2 \cdot CClF_2$; R13B1 = $CBrF_3$; R12 = $CCl_2F_2$; R22 = $CHClF_2$.

| Substance | Type | Exponent | Value | Ref. |
|---|---|---|---|---|
| $CO_2$ | u | $\alpha$ | 0.111 (1) | [50] |
|  |  | $\beta$ | 0.324 (2) |  |
| Cs | u | $\alpha$ | 0.13 (3) | [51] |
|  |  | $\beta$ | 0.355 (10) |  |
| Ethane | u | $\beta$ | 0.327 (2) | [52] |
|  |  | $\Delta$ | 0.46 (4) |  |
| Fluoroform | u | $\beta$ | 0.329 (1) | [53] |
| $GeH_4$ | u | $\beta$ | 0.333 (8) | [54] |
| $H_2$ | u | $\beta$ | 0.326 (3) | [55] |

| Substance | Type | Exponent | Value | Ref. |
|---|---|---|---|---|
| | | $\gamma$ | 1.19 (5) | |
| | | $\Delta$ | 0.46 (2) | |
| HD | u | $\beta$ | 0.352 (1) | [56] |
| HFC-32 | u | $\beta$ | 0.345 (1) | [57] |
| HFC-125 | u | $\beta$ | 0.341 (2) | [57] |
| Ne | u | $\beta$ | 0.3575 (10) | [56] |
| Ne, $N_2$ | u | $\beta$ | 0.327 (2) | [58] |
| | | $\Delta$ | 0.51 (3) | |
| | | $\Delta$ | 0.49 (5) | |
| R114 | u | $\beta$ | 0.312 | [59] |
| R13B1 | u | $\beta$ | 0.340 | [60] |
| R12 | u | $\beta$ | 0.337 | [61] |
| R22 | u | $\beta$ | 0.348 | [61] |
| Rb | u | $\alpha$ | 0.14 (3) | [51] |
| | | $\beta$ | 0.36 (1) | |
| $SF_6$ | u | $\beta$ | 0.48 (3) | [62] |
| | | $\beta$ | 0.325 (5) | [63] |
| | | $\beta$ | 0.338 | [64] |
| | | $\beta$ | 0.350 (4) | [65] |
| | | $\beta$ | 0.355 | [66] |
| | | $\gamma$ | 0.98 (5) | [62] |
| | | $\gamma$ | 1.24 (2) | [65] |
| | | $\delta$ | 3.0 (2) | [62] |
| 1,1,1,2-tetrafluoroethane | u | $\beta$ | 0.340 (1) | [67] |
| Xe | u | $\gamma$ | 1.246 (10) | [68] |
| Methanol + hexane | mx | $\gamma$ | 1.09 (3) | [69] |
| | | $\gamma - \alpha$ | 1.04(3) | |
| Methanol + cyclohexane | mx | $\beta$ | 0.33 (2) | [70] |
| | | $\gamma$ | 1.26 (5) | [71] |
| | | $\nu$ | 0.64 (2) | |
| Methanol + $n$-heptane | mx | $\beta$ | 0.3337 (5) | [72] |
| | | $\alpha$ | 0.11–0.35 | [73] |
| Methanol + isooctane | mx | $\beta$ | 0.323 (9) | [74] |
| $n$-hexane + $n$-tetradecafluorohexane | mx | $\beta$ | 0.35 (1) | [75] |
| Acetonitrile + cyclohexane | mx | $\beta$ | 0.322 (4) | [76] |
| Butylcellosolve + $H_2O$ | mx | $\alpha$ | 0.077 (41) | [77] |
| | | $\beta$ | 0.319 (14) | |
| | | $\gamma$ | 1.24 (1) | |
| | | $\nu$ | 0.606 (18) | |
| Iso-butoxyethanol + $H_2O$ | mx | $\alpha$ | 0.105 (8) | [78] |
| Isobutyric acid + $H_2O$ | mx | $\beta$ | 0.326 (3) | [79] |
| | | $\gamma$ | 1.19 (21) | [80] |
| | | $\nu$ | 0.654 (9) | |
| Deuterated cyclohexane + cyclohexane + $H_2O$ | mx | $\beta$ | 0.323 (3) | [81] |
| | | $\beta$ | 0.326 (2) | |
| | | $\beta$ | 0.322 (2) | |
| $n,n$-dimethylacetamide + octane | mx | $\beta$ | 0.324 (5) | [82] |
| | | $\beta$ | 0.329 (2) | |
| $n,n$-dimethylacetamide + decane | mx | $\beta$ | 0.329 (4) | [83] |
| | | $\beta$ | 0.333 (2) | |

| Substance | Type | Exponent | Value | Ref. |
|---|---|---|---|---|
| Ethylammoniumnitrate + $n$-octanol | mx | $\nu$ | 0.610 (6) | [84] |
| Ethyleneglycolmonoisobutylether + $H_2O$ | mx | $\beta$ | 0.332 (2) | [85] |
| 2,6-lutidine + $H_2O$ | mx | $\beta$ | 0.336 (30) | [86] |
| Nitroethane + cyclohexane, benzonitrile + isooctane | mx | $\beta$ | 0.325 (5) | [87] |
| Nitrobenzene + isooctane | mx | $\alpha$ | 0.145 (35) | [88] |
| Nitrobenzene + decane + benzene | mx | $\tilde{\beta}$ | 0.376 (8) | [89] |
| Perfluoroheptane + $CCl_4$ | mx | $\beta$ | 0.324 (5) | [90] |
| Triethylamine + $H_2O$ | mx | $\Delta$ | 0.52 (3) | [91] |
| Triethylamine + $H_2O$ + $D_2O$ | mx | $\alpha$ | 0.110 (4) | [92] |
| Trimethylethylammoniumbromide + chloroform | mx | $\nu$ | 0.621 (3) | [93] |
| Tetrachloromethane + tetradecafluoromethylcyclohexane | mx | $\beta$ | 0.289 (6) | [94] |
| Tetra-$n$-pentylammoniumbromide + $H_2O$ | mx | $\beta$ | 0.3370 (22) | [95] |
| | | $\beta$ | 0.3190 (11) | |
| | | $\beta$ | 0.3167 (16) | |
| $CO_2$ + $n$-butane | mx | $\beta$ | 0.359 | [96] |
| | | $\nu$ | 0.66 | |
| $CO_2$ + $n$-decane | mx | $\beta$ | 0.368 | [97] |
| | | $\nu$ | 0.646 | |
| Na + $NH_3$ | mx | $\beta$ | 0.34 (1) | [98] |
| | | $\Delta$ | 0.46 (3) | |
| Na + $ND_3$ | mx | $\gamma$ | 1.228 (39) | [99] |
| | | $\gamma$ | 1.2400 (157) | |
| | | $\gamma$ | 1.223 (19) | |
| | | $\eta$ | 0.0300 (15) | |
| | | $\eta$ | 0.0317 (13) | [100] |
| | | $\eta$ | 0.0302 (15) | |
| | | $\nu$ | 0.6279 (80) | [99] |
| $CoF_2$ | mg | $\alpha$ | 0.109 (6) | [101] |
| DyAlG | mg | $\beta$ | 0.33 (1) | [102] |
| $FeCl_2$ | mg | $\alpha$ | 0.15(4) | [103] |
| $FeF_2$ | mg | $\alpha$ | 0.111 (7) | [104] |
| | | $\alpha$ | 0.115 (4) | |
| | | $\alpha$ | 0.11 (3) | [105] |
| | | $\beta$ | 0.325 (2) | [106] |
| | | $\gamma$ | 1.25 (1) | [107] |
| $MnBr_2$ | mg | $\alpha$ | 0.118 (7) | [108] |
| $MnCl_2$ | mg | $\beta$ | 0.297 (3) | [109] |
| $MnF_2$ | mg | $\alpha$ | 0.123 (5) | [104] |
| | | $\alpha$ | 0.091 (5) | |
| $NdRu_2Si_2$ | mg | $\alpha$ | 0.11 (3) | [110] |
| $U_3P_4$ | mg | $\beta$ | 0.315 (15) | [111] |
| | | $\beta$ | 0.313 (15) | |
| | | $\gamma$ | 1.25 (2) | |
| AOT + $n$-decane + $H_2O$ | mi | $\gamma$ | 1.26 (10) | [112] |
| | | $\gamma$ | 1.22 (5) | [113] |
| | | $\gamma$ | 1.61 (9) | [114] |
| | | $\nu$ | 0.61 (6) | [112] |
| | | $\nu$ | 0.75 (5) | [113] |
| | | $\nu$ | 0.68 (8) | |
| | | $\nu$ | 0.72 (4) | [114] |

| Substance | Type | Exponent | Value | Ref. |
|---|---|---|---|---|
| Benzene + BHDC + $H_2O$ | mi | $\beta$ | 0.34 (8) | [115] |
| | | $\gamma$ | 1.18 (3) | [116] |
| | | $\nu$ | 0.60 (2) | [116] |
| 2-butoxyethanol + $D_2O$ | mi | $\gamma$ | 1.216 (13) | [117] |
| | | $\eta$ | 0.039 (4) | |
| | | $\nu$ | 0.623 (13) | |
| $C_6E_3$ + $H_2O$ | mi | $\alpha$ | 0.11 (4) | [118] |
| | | $\beta$ | 0.327 (4) | |
| | | $\gamma$ | 1.24 (1) | |
| | | $\gamma$ | 1.241 (16) | [119] |
| | | $\eta$ | 0.016 (5) | [118] |
| | | $\nu$ | 0.627 (6) | |
| | | $\nu$ | 0.632 (11) | [119] |
| $C_8E_4$ + $H_2O$ | mi | $\gamma$ | 1.237 (7) | [119] |
| | | $\gamma$ | 1.243 (7) | |
| | | $\nu$ | 0.630 (12) | |
| | | $\nu$ | 0.630 (18) | |
| $C_{10}E_4$ + $H_2O$ | mi | $\gamma$ | 1.25 (2) | [120] |
| | | $\nu$ | 0.63 (1) | |
| $C_{12}E_5$ + $H_2O$ | mi | $\gamma$ | 1.17 (11) | [121] |
| | | $\nu$ | 0.65 (4) | |
| $C_{12}E_6$ + $D_2O$ | mi | $\gamma$ | 1.2 (1) | [122] |
| | | $\nu$ | 0.60 (5) | |
| $C_{12}E_6$ + $H_2O$ | mi | $\gamma$ | 1.2 (1) | [123] |
| | | $\nu$ | 0.60 (3) | |
| $C_{12}E_8$ + $D_2O$ | mi | $\gamma$ | 1.21 (2) | [124] |
| | | $\nu$ | 0.62 (2) | |
| $C_{12}E_8$ + $H_2O$ | mi | $\gamma$ | 1.20 (4) | [124] |
| | | $\nu$ | 0.63 (1) | |
| Sodiumdodecylsulfate + butanol + NaCl | mi | $\nu$ | 0.62 (3) | [125] |
| | | $\nu$ | 0.64 (4) | |
| | | $\nu$ | 0.63 (5) | |
| Cationic surfactant in aqueous salt solution | mi | $\tilde{\beta}$ | 0.375 (10) | [126] |
| | | $\tilde{\gamma}$ | 1.39 (4) | |
| | | $\tilde{\nu}$ | 0.70 (3) | |

# References

[1] G.A. Baker, B.G. Nickel and D.I. Meiron, Phys. Rev. B **17**, 1365 (1978).

[2] J.C. Le Guillou and J. Zinn-Justin, Phys. Rev. B **21**, 3976 (1980).

[3] J. Adler, J. Phys. A **16**, 3585 (1983).

[4] J.C. Le Guillou and J. Zinn-Justin, J. Physique Lett. **46**, L137 (1985).

[5] J.C. Le Guillou and J. Zinn-Justin, J. Phys. (Paris) **48**, 19 (1987).

[6] A.J. Liu and M.E. Fisher, Physica A **156**, 35 (1989).

[7] B.G. Nickel and J.J. Rehr, J. Stat. Phys. **61**, 1 (1990).

[8] B.G. Nickel, Physica A **177**, 189 (1991).

[9] A.J. Guttmann and I.G. Enting, J. Phys. A **27**, 8007 (1994).

[10] M. Kolesik and M. Suzuki, Physica A **215**, 138 (1995).

[11] H.W.J. Blöte and R.H. Swendsen, Phys. Rev. B **20**, 2077 (1979).

[12] G.S. Pawley, R.H. Swendsen, D.J. Wallace and K.G. Wilson, Phys. Rev. B **29**, 4030 (1984).

[13] H.W.J. Blöte, J.A. de Bruin, A. Compagner, J.H. Croockewit, Y.T.J.C. Fonk, J.R. Heringa, A. Hoogland and A.L. van Willigen, Europhys. Lett. **10**, 105 (1989).

[14] H.W.J. Blöte, A. Compagner, J.H. Croockewit, Y.T.J.C. Fonk, J.R. Heringa, A. Hoogland, T.S. Smit and A.L. van Willigen, Physica A **161**, 1 (1989).

[15] N.A. Alves, B.A. Berg and R. Villanova, Phys. Rev. B **41**, 383 (1990).

[16] N. Ito and M. Suzuki, J. Phys. Soc. Jpn. **60**, 1978 (1991).

[17] A.M. Ferrenberg and D.P. Landau, Phys. Rev. B **44**, 5081 (1991).

[18] C.F. Baillie, R. Gupta, K.A. Hawick and G.S. Pawley, Phys. Rev. B **45**, 10438 (1992).

[19] H.W.J. Blöte and G. Kamieniarz, Physica A **196**, 455 (1993).

[20] D.P. Landau, Physica A **205**, 41 (1994).

[21] E.W. Meyer, Master's thesis, Delft University, 1991, unpublished.

[22] A. Hoogland, J. Spaa, B. Selman and A. Compagner, J. Comp. Phys. **51**, 250 (1983).

[23] A. Hoogland, A. Compagner and H.W.J. Blöte, *The Delft Ising System Processor*, Chapter 7 of *Architecture and Performance of Specialized Computers*, ed. B. Alder, in the series *Computational Techniques* (Academic, New York, 1988).

[24] R.H. Swendsen and J.S. Wang, Phys. Rev. Lett. **58**, 86 (1987).

[25] C.F. Baillie and P.D. Coddington, Phys. Rev. B **43**, 10617 (1991).

[26] U. Wolff, Phys. Rev. Lett. **60**, 1461 (1988).

[27] M.N. Barber, R.B. Pearson, D. Toussaint and J.L. Richardson, Phys. Rev. B **32**, 1720 (1985).

[28] G. Parisi and F. Rapuano, Phys. Lett. B **157**, 301 (1985).

[29] A. Hoogland, A. Compagner and H.W.J. Blöte, Physica A **132**, 593 (1985).

[30] G. Bhanot, D. Duke and R. Salvador, Phys. Rev. B **33**, 7841 (1986).

[31] A.M. Ferrenberg, D.P. Landau and Y.J. Wong, Phys. Rev. Lett. **69**, 3382 (1992).

[32] T.G. Lewis and W.H. Payne, J. Ass. Comp. Mach. **20**, 456 (1973).

[33] W. Selke, A.L. Talapov and L.N. Shchur, Pis'ma Zh. Eksp. Teor. Fiz. **58**, 684 (1993).

[34] K. Binder, Z. Phys. B **43**, 119 (1981).

[35] F. Livet, Europhys. Lett. **16**, 139 (1991).

[36] F. Livet, private communication (1993).

[37] L.N. Shchur and H.W.J. Blöte, to be published (1995).

[38] J.R. Heringa, H.W.J. Blöte and A. Compagner, Int. J. Mod. Phys. C **3**, 561 (1992).

[39] L.N. Shchur and H.W.J. Blöte, unpublished results (1995).

[40] R.B. Pearson, Phys. Rev. B **26**, 6285 (1982).

[41] H. Saleur and B. Derrida, J. Phys. (Paris) **46**, 1043 (1985).

[42] M. Suzuki, Prog. Theor. Phys. **58**, 1142 (1977).

[43] E. Brézin, J. Phys. (Paris) **43**, 15 (1982).

[44] V. Privman, P.C. Hohenberg and A. Aharony, *Universal Critical-Point Amplitude Relations*, in: *Phase Transitions and Critical Phenomena*, eds. C. Domb and J.L. Lebowitz (Academic Press, London, 1991), Vol. 14, pp. 1–134.

[45] P.-Y. Lai and K.K. Mon, Phys. Rev. B **40**, 11120 (1989).

[46] G.A. Baker, Jr., and N. Kawashima, Phys. Rev. Lett. **75**, 994 (1995).

[47] A. Rosengren, J. Phys. A **19**, 1709 (1986).

[48] A. Kumar, H.R. Krishnamurthy and E.S.R. Gopal, Phys. Rep. **98**, 57 (1983).

[49] M.E. Fisher, Phys. Rev. **176**, 257 (1968).

[50] I.M. Abdulagatov, N.G. Polikhsonidi and R.G. Batyrova, J. Chem. Therm. **26**, 1031 (1994).

[51] S. Jüngst, B. Knuth and F. Hensel, Phys. Rev. Lett. **55**, 2160 (1985).

[52] J.R. de Bruyn and D.A. Balzarini, Phys. Rev. A **36**, 5677 (1987).

[53] U. Närger, J.R. de Bruyn, M. Stein and D.A. Balzarini, Phys. Rev. B **39**, 11914 (1989).

[54] D. Balzarini, O.G. Mouritsen and P. Palffy-Muhoray, Can. J. Phys. **61**, 1301 (1983).

[55] J.R. de Bruyn and D.A. Balzarini, Phys. Rev. B **39**, 9243 (1989).

[56] E.T. Shimanskaya, A.V. Oleinikova and Yu.I. Shimanskii, Sov. J. Low Temp. Phys. **16**, 780 (1990).

[57] S. Kuwabara, H. Aoyama, H. Sato and K. Watanabe, J. Chem. Eng. Data **40**, 112 (1995).

[58] M.W. Pestak and M.H.W. Chan, Phys. Rev. B **30**, 274 (1984).

[59] Y. Higashi, M. Uematsu and K. Watanabe, Bull. JSME **28**, 2968 (1985).

[60] Y. Higashi, M. Uematsu and K. Watanabe, Bull. JSME **28**, 2660 (1985).

[61] Y. Higashi, S. Okazaki, Y. Takaishi, M. Uematsu and K. Watanabe, J. Chem. Eng. Data **29**, 31 (1984).

[62] W. Wagner, N. Kurzeja and B. Pieperbeck, Fluid Phase Eq. **79**, 151 (1992).

[63] B.J. Thijsse, J. Chem. Phys. **74**, 4678 (1981).

[64] S.N. Biswas and C.A. ten Seldam, Fluid Phase Eq. **47**, 67 (1989).

[65] K. Morofuji, K. Fujii, M. Uematsu and K. Watanabe, Int. J. Thermophys. **7**, 17 (1986).

[66] A.C. Michels, Ber. des 18. Bunsen-Kolloquium der Bunsenges. Phys. Chem. 53 (1983).

[67] Y. Kabata, S. Tanokawa, M. Uematsu and K. Watanabe, Int. J. Thermophys. **10**, 605 (1989).

[68] H. Güttinger and D.S. Cannell, Phys. Rev. A **24**, 3188 (1981).

[69] S. Kawase, K. Maruyama, S. Tamaki and H. Okazaki, J. Phys. Cond. Matt. **6**, 10237 (1994).

[70] R.R. Singh and W.A. van Hook, J. Chem. Phys. **87**, 6097 (1987).

[71] D.T. Jacobs, Phys. Rev. A **33**, 2605 (1986).

[72] A.G. Aizpiri, J.A. Correa, R.G. Rubio and M. Diaz Peña, Phys. Rev. B **41**, 9003 (1990).

[73] J. Balakrishnan, M.K. Gunasekaran and E.S.R. Gopal, Ind. J. Pure Appl. Phys. **22**, 286 (1984).

[74] A.C. Ploplis, P.S. Wardwell and D.T. Jacobs, J. Phys. Chem. **90**, 4676 (1986).

[75] G.I. Pozharskaya, N.L. Kasapova, V.P. Skripov and Yu.D. Kolpakov, J. Chem. Therm. **16**, 267 (1984).

[76] V. Vani, S. Guha and E.S.R. Gopal, J. Chem. Phys. **84**, 3999 (1986).

[77] K. Hamano, T. Kawazura, T. Koyama and N. Kuwahara, J. Chem. Phys. **82**, 2718 (1985).

[78] U. Würz, M. Grubić and D. Woermann, Ber. Bunsenges. Phys. Chem. **96**, 1460 (1992).

[79] W.V. Andrew, T.B.K. Khoo and D.T. Jacobs, J. Chem. Phys. **85**, 3985 (1986).

[80] L.W. DaMore and D.T. Jacobs, J. Chem. Phys. **97**, 464 (1992).

[81] C. Houessou, P. Guenoun, R. Gastaud, F. Perrot and D. Beysens, Phys. Rev. A **32**, 1818 (1985).

[82] X. An and W. Shen, J. Chem. Therm. **26**, 461 (1994).

[83] X. An, W. Shen, H. Wang and G. Zheng, J. Chem. Therm. **25**, 1373 (1994).

[84] W. Schröer, S. Wiegand and H. Weingärtner, Ber. Bunsenges. Phys. Chem. **97**, 975 (1993).

[85] M. Nakata, T. Dobashi, N. Kuwahara and M. Kaneko, J. Chem. Soc. Far. Trans. II **78**, 1801 (1982).

[86] V. Balevicius, N. Weiden and A. Weiss, Z. Naturforsch. A **47**, 583 (1992).

[87] R. Kindt, J. Thoen and W. Van Dael, Int. J. Thermophys. **9**, 749 (1988).

[88] M. Merabet and T.K. Bose, Phys. Rev. A **25**, 2281 (1982).

[89] S.J. Rzoska and J. Chrapeć, J. Chem. Phys. **90**, 2783 (1989).

[90] D.T. Jacobs, J. Phys. Chem. **86**, 1895 (1982).

[91] A. Bourgou and D. Beysens, Phys. Rev. Lett. **47**, 257 (1981).

[92] E. Bloemen, J. Thoen and W. Van Dael, J. Chem. Phys. **75**, 1488 (1981).

[93] S. Wiegand, M. Kleemeier, J.-M. Schröder, W. Schröer and H. Weingartner, Int. J. Thermophys. **15**, 1045 (1994).

[94] M.K. Kumaran, C.J. Halpin and G.C. Benson, J. Chem. Therm. **15**, 1071 (1983).

[95] M.L. Japas and J.M.H. Levelt-Sengers, J. Phys. Chem. **94**, 5361 (1990).

[96] J. J.-C. Hsu, N. Nagarajan and R.L. Robinson, J. Chem. Eng. Data **30**, 485 (1985).

[97] N. Nagarajan and R.L. Robinson, J. Chem. Eng. Data **31**, 168 (1986).

[98] B.K. Das and S.C. Greer, J. Chem. Phys. **74**, 3630 (1981).

[99] P. Damay, F. Leclercq and P. Chieux, Phys. Rev. B **40**, 4696 (1989).

[100] P. Damay, F. Leclercq and P. Chieux, Physica B **156**, 223 (1989).

[101] C.A. Ramos, A.R. King and V. Jaccarino, Phys. Rev. B **40**, 7124 (1989).

[102] J.M. Hastings, L.M. Corliss and W. Kunnmann, Phys. Rev. B **31**, 2902 (1985).

[103] J. Kushauer and W. Kleemann, J. Phys. Cond. Matt. **7**, L1 (1995).

[104] D.P. Belanger, P. Nordblad, A.R. King, V. Jaccarino, L. Lundgren and O. Beckman, J. Magn. Magn. Mater. **31–34**, 1095 (1983).

[105] M. Marinelli, F. Mercuri and D.P. Belanger, J. Magn. Magn. Mater. **140–144**, 1547 (1995).

[106] J. Mattsson, C. Djurberg and P. Nordblad, J. Magn. Magn. Mater. **136**, L23 (1994).

[107] D.P. Belanger and H. Yoshizawa, Phys. Rev. B **35**, 4823 (1987).

[108] S.N. Bhatia, Pramãna **18**, 249 (1982).

[109] J.M. Steijger, E. Frikkee, L.J. de Jongh and W.J. Huiskamp, J. Magn. Magn. Mater. **31–34**, 1091 (1983).

[110] M.A. Salgueiro, B.G. Almeida, M.M. Amado, J.B. Sousa, B. Chevalier and J. Étourneau, J. Magn. Magn. Mater. **125**, 103 (1993).

[111] A.M. Strydom, P. de V. du Plessis, D. Kaczorowski and R. Troć, Physica B **186–188**, 785 (1993).

[112] P. Honorat, D. Roux and A.M. Bellocq, J. Physique Lett. **45**, L961 (1984).

[113] J.S. Huang and M.W. Kim, Phys. Rev. Lett. **47**, 1462 (1981).

[114] M. Kotlarchyk, S.-H. Chen and J.S. Huang, Phys. Rev. A **28**, 508 (1983).

[115] R. Aschauer and D. Beysens, J. Chem. Phys. **98**, 8194 (1993).

[116] R. Aschauer and D. Beysens, Phys. Rev. E **47**, 1850 (1993).

[117] J. Schmitz, L. Belkoura and D. Woermann, Ann. Phys. (Leipzig) **3**, 1 (1994).

[118] K. Hamano, T. Kaneko, K. Fukuhara and N. Kuwahara, Int. J. Thermophys. **10**, 389 (1989).

[119] A. Zielesny, L. Belkoura and D. Woermann, Ber. Bunsenges. Phys. Chem. **98**, 579 (1994).

[120] K. Hamano, N. Kuwahara, I. Mitsushima, K. Kubota and T. Kamura, J. Chem. Phys. **94**, 2172 (1991).
[121] C. Sinn and D. Woermann, Ber. Bunsenges. Phys. Chem. **96**, 913 (1992).
[122] J.P. Wilcoxon, D.W. Schaefer and E.W. Kaler, J. Chem. Phys. **90**, 1909 (1989).
[123] J.P. Wilcoxon and E.W. Kaler, J. Chem. Phys. **86**, 4684 (1987).
[124] G. Dietler and D.S. Cannell, Phys. Rev. Lett. **60**, 1852 (1988).
[125] O. Abillon, D. Chatenay, D. Langevin and J. Meunier, J. Physique Lett. **45**, L223 (1984).
[126] K. Kubota, N. Kuwahara and H. Sato, J. Chem. Phys. **100**, 4543 (1994).